\documentclass[%
 reprint,
superscriptaddress,
nofootinbib,
 amsmath,amssymb,
 aps,prd,
]{revtex4-2}
\usepackage{amsmath}
\usepackage[section]{placeins}
\usepackage[utf8]{inputenc}
\usepackage[english]{babel}
\usepackage{float}
\usepackage{isotope}

\usepackage[dvipsnames]{xcolor}

\usepackage{soul}
\usepackage{graphicx}
\usepackage{dcolumn}
\usepackage{bm}

\usepackage{graphicx}
\usepackage{dcolumn}
\usepackage{bm}

\usepackage[draft, inline, nomargin]{fixme}
\fxsetup{theme=color, mode=multiuser}

\FXRegisterAuthor{bc}{1}{\color{red}BC}
\FXRegisterAuthor{ag}{2}{\color{blue}AG}
\FXRegisterAuthor{ph}{3}{\color{magenta}PH}

\begin{document}


\title{The use of CEvNS to monitor spent nuclear fuel}

\author{Caroline von Raesfeld}
\email{carolinevr@g.ucla.edu}
\affiliation{Department of Physics and Astronomy, UCLA, CA, 90024}
\affiliation{Center for Neutrino Physics, Physics Department, Virginia Tech, Blacksburg, VA 24061}

\author{Patrick Huber}
\email{pahuber@vt.edu}
\affiliation{Center for Neutrino Physics, Physics Department, Virginia Tech, Blacksburg, VA 24061}

\date{\today}%

\begin{abstract}

Increasing amounts of spent nuclear fuel are stored in dry storage casks for prolonged periods of time. To date no effective technology exists to re-verify cask contents should this become necessary. We explore the applicability of Coherent-Elastic Neutrino-Nucleus Scattering (CEvNS) to monitor the content of spent nuclear fuel (SNF) from dry storage casks. SNF produces neutrinos chiefly from  $^{90}$Sr decays. We compare these results with what can be achieved via Inverse Beta Decay (IBD). We demonstrate that at low nuclear recoil energies CEvNS events rates exceed the IBD event rates by 2--3 orders of magnitude for a given detector mass. We find that  a 10\,kg argon or germanium detector 3 meters from a fuel cask can detect over 100 events per year if a nuclear recoil threshold under 100 eV can be achieved.
\end{abstract}

\maketitle

\section{\label{sec:introduction}Introduction}

Since  Fermi built the first nuclear reactor, the question of what to do with spent nuclear fuel (SNF) has been difficult to answer. Most nations plan to eventually put their spent fuel into long-term geological storage. However, so far only Finland~\cite{onkalo} and Sweden~\cite{forsmark} have specific plans and facilities to this end, with operation start dates in the next decade. For the first few years after discharge from a reactor, fuel is typically put into water pools (wet storage), which provide both radiation shielding and cooling. The nuclear accident at Fukushima Daichi in the aftermath of the 2011 tsunami has made it all too obvious that wet storage is not a safe solution for longer periods of time and has re-emphasized the critical importance to move SNF into much safer dry storage as soon as possible. This lesson combined with the lack of geological repositories and the limited capacity of wet storage facilities results in ever increasing amounts of SNF in dry storage for ever longer periods, in many cases exceeding decades. In the U.S. alone 80\,000 tons of spent fuel are held in dry storage and each year approximately 2\,000 tons are added~\cite{snf}. In dry storage, 10-20 spent fuel assemblies are put together in a gas-tight steel cask, which in turn is put inside a concrete shell for added radiation protection and to protect the steel from the elements. To safeguard the spent fuel, the initial inventory of the cask at loading is verified and seals are then applied to the cask. Verifying the integrity of these seals requires inspectors to climb on top of the cask which incurs the risk of falls and radiation exposure. Moreover, these inspections are physically demanding and time consuming. Should a seal fail, a re-verification of the cask content becomes necessary, ideally by some means of non-destructive essay. However, the cask produces relatively small and unspecific radiation signatures due to the fuel itself having a density of around 10\,g$\,\mathrm{cm}^{-3}$, which results in severe self-shielding on top of the fact that the whole purpose of the cask is to shield radiation. As a result, no conventional technique based on either neutron or gamma emission has been found satisfactory to verify a cask's fuel content~\cite{nutools}. The other, rather cumbersome, option is to bring the cask back to a spent fuel pond and to re-open it, which incurs the attendant risk of radiation exposure to inspectors.  This motivates research into other methods of fuel verification, and several recent developments have positioned neutrinos as an interesting option.

The use of neutrinos for applications in reactor monitoring in a nuclear security context has been widely discussed, for a recent review see Ref.~\cite{Bernstein:2019hix} and references therein. The main advantage of neutrinos compared to say, ionizing radiation or neutron signatures, is the fact that they can penetrate arbitrary amounts of material and thus, can "see" into places like the core of a running nuclear reactor which are otherwise inaccessible. This in turn allows placement of detectors outside of the reactor building or even of the facility, rendering this technique non-intrusive. At the same time, neutrino measurements provide a direct indication of the reactor core inventory without reference to the prior operating of fueling history. Most of the previous literature explores neutrino\footnote{A nuclear reactor, or beta decay in general, only produces electron antineutrinos, which we will  simply call neutrinos.} detection via inverse beta decay (IBD). Studies of coherent elastic neutrino nucleus scattering (CEvNS) in a reactor monitoring context are still relatively rare, see for instance Refs.~\cite{Cogswell:2016aog,Bowen:2020unj,Cogswell:2021qlq}. The specific application of neutrino monitoring to spent nuclear fuel both in dry storage casks and geological repositories based on IBD has been studied previously~\cite{Brdar2017}. In monitoring SNF, neutrinos offer the advantage that they are not afflicted by self-shielding of the fuel or attenuation by the cask. The bulk of neutrino emission in SNF older than a few years stems from the $^{90}$Sr/$^{90}$Y decay chain, which remains sizeable even after many decades due to the  $^{90}$Sr's half life of 28 years.  In the recent NuTools study~\cite{nutools} which is focused on neutrino applications in nuclear security and energy and has been charted by the U.S. National Nuclear Security Administration, monitoring of spent fuel in dry storage casks has been identified as a promising field for further study for the reasons outlined here.

\section{\label{sec:flux} Neutrino Flux from Spent Nuclear Fuel}

Neutrino fluxes from SNF are less prominent than fluxes from active reactors, and thus detection of neutrino events using SNF fluxes is more technologically challenging. At the point that SNF is transferred to dry storage casks, typically several years after discharge from the reactor, the radioactivity comes from long-lived fission products. After several years nearly all neutrino emission stems from from $^{90}$Sr/Y.
    
In 2017, Brdar \textit{et al.} explored the applicability of using Inverse Beta Decay (IBD) to monitor the amount of SNF  in dry storage casks~\cite{Brdar2017}. In this study, the neutrino flux from SNF was calculated as a function of time since its discharge from a nominal light water pressurized water reactor corresponding to a fuel burnup of 45\,GW day per ton; we will use these fluxes in our analysis as well.  These fluxes were then used to calculate how many IBD events could be observed for a variety of IBD detector setups.  It was demonstrated that such neutrino detectors could be useful to remotely detect any changes in the SNF content with detector masses in the 10s of ton range. Shortly thereafter, Coherent Elastic Neutrino Nucleus Scattering (CEvNS) was discovered by the COHERENT collaboration~\cite{Akimov:2017ade} using neutrinos from Spallation Neutron Sources (SNS) at Oak Ridge National Laboratory with a mean energy of around 30\,MeV. In the CEvNS reaction, the detectable signature arises solely from nuclear recoil, which requires detectors of exquisite low-energy sensitivity and careful background mitigation. The relatively high neutrino energy paired with pulsed nature of the SNS neutrino source  allowed the initial detection to succeed. Detecting neutrinos from a reactor via CEvNS is more challenging since the mean neutrino energy is about an order of magnitude smaller and hence the recoil energies are much lower. Additionally, reactors are continuous neutrino sources. There are a significant number of collaborations currently attempting reactor CEvNS detection, including CONUS~\cite{Bonet:2020awv}, MINER~\cite{Agnolet:2016zir}, 
NUCLEUS~\cite{Rothe:2019aii}, RED-100~\cite{Akimov:2019ogx},  NEON~\cite{Choi:2020gkm}, RICOCHET~\cite{Billard:2016giu}, TEXONO~\cite{Wong:2016lmb}, CONNIE~\cite{Aguilar-Arevalo:2019jlr}. There are numerous goals in basic science related to reactor CEvNS~\cite{deGouvea:2020hfl,Fernandez-Moroni:2020yyl,Miranda:2020zji,Miranda:2020syh,Tomalak:2020zfh} as well as the aforementioned potential applications to reactor monitoring for nuclear security~\cite{Cogswell:2016aog,Bernstein:2019hix,Bowen:2020unj,Cogswell:2021qlq}.

These efforts make it appear worthwhile to explore how CEvNS might  apply to SNF monitoring.

\section{\label{sec:reactions} CEvNS and IBD reactions}
A criterion of interest in this study is to expand on the results found in Brdar \textit{et al.} by quantitatively understanding any advantage a CEvNS detector might have over an IBD detector. Each reaction has been well-documented and are now presented here.

In Inverse Beta Decay (IBD) an electron antineutrino interacts with a proton, resulting in  a neutron and a positron.
\begin{equation}
    \bar{\nu}_e + p \rightarrow n + e^+
\end{equation}
The signal seen by an IBD detector arrives in two parts, first is a prompt energy deposition from the positron and after a slight delay another energy deposition will be produced by the neutron as it undergoes neutron capture. Using timing and spatial localization the two signals can be correlated in a delayed coincidence. The IBD reaction has a neutrino energy threshold of $m_n-m_p+2 m_e \simeq 1.8\,\text{ MeV}.$
IBD and delayed coincidence were used in the discovery of neutrinos~\cite{Cowan:1956rrn} and have been used for all reactor neutrino detection experiments since then. We use the IBD cross section from Ref.~\cite{Vogel1999} and assume a detector chemical makeup of CH$_2$.

Coherent-Elastic Neutrino-Nucleus Scattering  occurs when a neutrino of any flavor interacts elastically with a nucleus, producing a nuclear recoil signal~\cite{Freedman:1973yd}
\begin{equation}
    \bar{\nu} + N \rightarrow \bar{\nu} + N
\end{equation}
This reaction has a kinematic limit for the nuclear recoil energy, given by
\begin{equation}
T_{max} = \frac{E_\nu}{1 + \frac{M_N}{2E_\nu}}
\end{equation}
The nuclear recoil is dependent on the incident neutrino energy, and thus occurs at relatively low energies. Due to this, CEvNS long evaded detection even though it was postulated in 1974~\cite{Freedman:1973yd}. The relevant CEvNS cross section is given as
\begin{equation}
\frac{d\sigma_{CEV}}{dT} = \frac{G_f^2}{4\pi} N_N ^2 M_N \left( 1 -\frac{ M_N \
T}{2 E_\nu ^2}\right) 
\end{equation}
where $G_f$ is the Fermi constant, $N_N$ is the target isotope's neutron number, $M_N$ is the mass of a nucleus of the target isotope, $E_\nu$ is the incident neutrino energy and $T$ is the nuclear recoil energy.

\section{\label{sec:event rates}comparison of IBD and CEvNS Event Rates}

\begin{figure}[ht]
    \centering
    \includegraphics[width = \columnwidth]{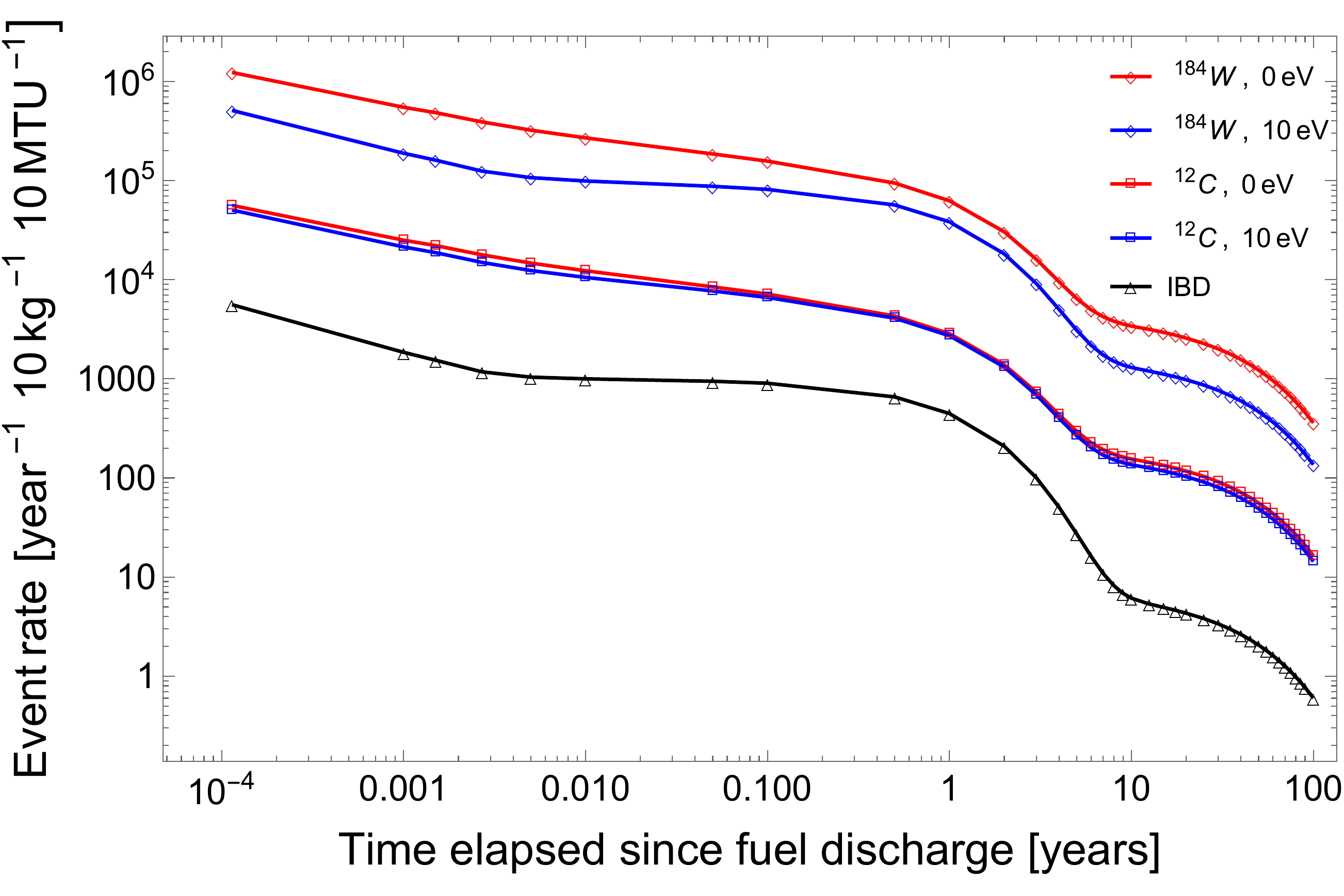}
    \caption{ Event rate of CEvNS for a \isotope[184]{W} and \isotope[12]{C} detector able to resolve 0 eV [red] or 10 eV [blue], compared to the IBD event rate of a detector the same size. The detector mass is 10\,kg, the standoff is 3\,m and data taking period is one year from a 10\,MTU source.}
    \label{fig:tungsten carbon ratios}
\end{figure}

Several factors are usually cited to favor a CEvNS detector to have an advantage over an IBD detector in this context. Firstly, the $N^2$ dependence of the CEvNS cross section allows for much larger cross sections than IBD, emphasized for large isotopes with high neutron numbers. Secondly, IBD is limited by the 1.8\,MeV neutrino energy threshold, whereas CEvNS can occur at any energy and thus can probe into the neutrino fluxes below the IBD threshold. It has been shown in the context of reactor neutrinos that these two factors do not provide a decisive advantage~\cite{Bowen:2020unj}. We find that in the context of SNF the real advantage arises from phase space: the Q-value of $^{90}$Y beta decay is only  2.28\,MeV~\cite{ensdf} compared to the IBD threshold of 1.8\,MeV, which results in an effective IBD cross section of only $8\times10^{-45}\,\mathrm{cm}^2$. CEvNS being a threshold-less reaction can access a much larger fraction of the available phase space resulting in a very much enhanced cross section. This is illustrated in Fig.~\ref{fig:tungsten carbon ratios}, where we compare the event rates in a 10\,kg detector 3\,m from 10\,MTU\footnote{Metal Tons of Uranium (MTU) is a common unit for nuclear fuel. 10 MTU roughly corresponds to the contents of one dry storage cask.} of SNF for one year of exposure.   CEvNS event rates from isotopes \isotope[12]{C} and \isotope[184]{W} and IBD event rates are shown as a function of the time since the discharge of the spent fuel. These two isotopes are chosen because they bracket plausible detector materials in terms of the neutron number $N_N$.
 
A key parameter of interest in this evaluation is the nuclear recoil energy that a CEvNS detector would need to be able to resolve. Many running or in-progress CEvNS detectors aim to resolve nuclear recoil energies below 100\,eV, see for instance~\cite{Agnolet:2016zir,Rothe:2019aii,Billard:2016giu,Aguilar-Arevalo:2019jlr},
thus a range of 0--100\,eV was considered for the following analyses. For the event rates shown in Fig.~\ref{fig:tungsten carbon ratios}, it was assumed that the CEvNS detector would be able to resolve either all nuclear recoil energies (0\,eV)  or down to 10\,eV. These low values may be out of the range of feasibility of current detectors, but it displays, effectively, the maximum advantage that can be achieved with a CEvNS detector. 

For \isotope[184]{W} with a large cross section, the advantage reaches approximately three orders of magnitude over the IBD event rate. Even for \isotope[12]{C} with a smaller cross section the advantage is more than one  order of magnitude.  After an elapsed time of around 10 years, the event rates in each case including IBD have the same time dependence. Most spent fuel contained in dry storage casks is 10-70 years of age, thus the relative event rate between CEvNS and IBD is not expected to change with time.
In further analyses an average discharge time of 30 years for the fuel is used.

Figure~\ref{fig:tungsten carbon ratios} also shows that while \isotope[184]{W} produces the highest event rate for 0\,eV nuclear energy threshold, an increase to 10\,eV will cause the event rate to decrease significantly. In contrast, the event rate of \isotope[12]{C} stays comparably stable through the increase to 10\,eV. This indicates a trade off between the $N^2$ dependence of the cross section and the maximum nuclear recoil energy $T_{max}$, which effectively scales as $\frac{1}{M}$, where $M$ is the isotope mass. Thus higher mass isotopes have a larger cross section, but have much lower maximum recoil energies. This acts as a limiting factor, asserting that a detector must be able to resolve very low recoil energies to obtain these high event rates coming from heavier isotopes. 

\begin{figure}[ht]
    \centering
    \includegraphics[width = \columnwidth]{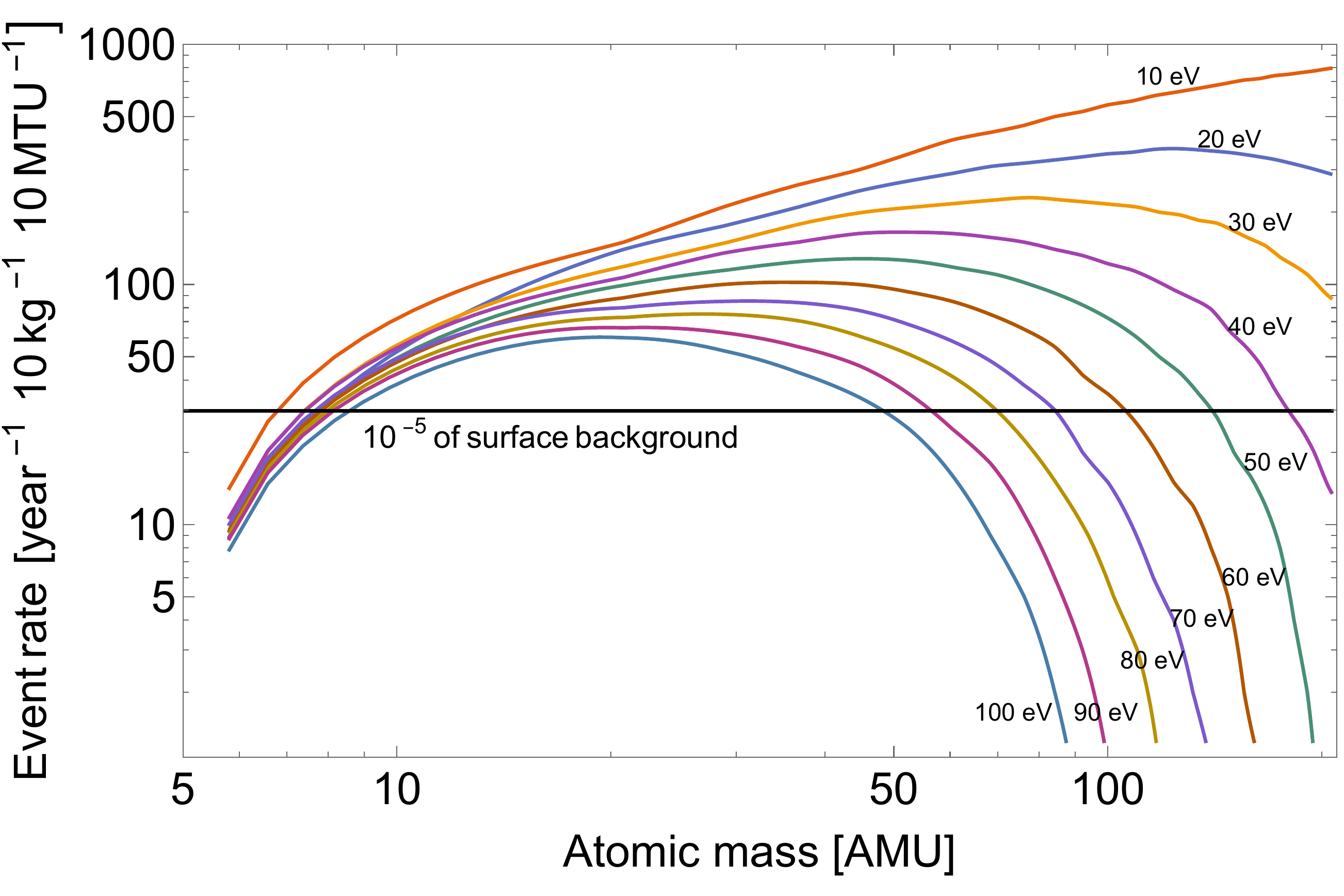}
    \caption{CEvNS event rate as a function of the atomic mass of an isotope, with labeled nuclear recoil thresholds. High mass isotopes have high event rates at low recoil thresholds and at high thresholds low mass isotopes perform best. The detector mass is 10\,kg, the standoff is 3\,m and data taking period is one year from a 10\,MTU source.}
    \label{fig:isotopes}
\end{figure}

The relationship between atomic mass, the resolvable nuclear recoil energy, and the resulting event rate is shown in Fig.~\ref{fig:isotopes}. For lower nuclear recoil thresholds like 10 or 20\,eV, the event rates increase as atomic mass increases, but at energies of even 30\,eV it is apparent that there is a maximum event rate as the line begins to drop off at higher masses. These become steeper with higher resolvable recoil energies. For reference, the approximate background rate corresponding to $10^{-5}$ of surface level cosmic ray neutrons is shown.

\section{\label{sec:mle}Verification of Cask Contents}

The ultimate goal of the CEvNS detector is to verify the fuel content in a dry storage cask, to account for any otherwise undetectable losses in fuel. A key limitation will arise from backgrounds induced by cosmic rays, in particular from neutrons which result in nuclear recoils which are indistinguishable from CEvNS events. The actual background levels with moderate shielding close to the surface are an active area of study and overall not well known~\cite{excess} and thus we will show the following results a function of the recoil background, where values of $10^4$ or larger correspond roughly to the rates without shielding at the surface. The task is to accurately measure the true amount of fuel in a  dry storage cask. Assuming a true mass of 10\,MTU,  a maximum likelihood estimate was carried out to explore how well a potential detector with the aforementioned ideal parameters could measure the true fuel content.

\begin{figure}[htb]
    \centering
\includegraphics[width =\columnwidth]{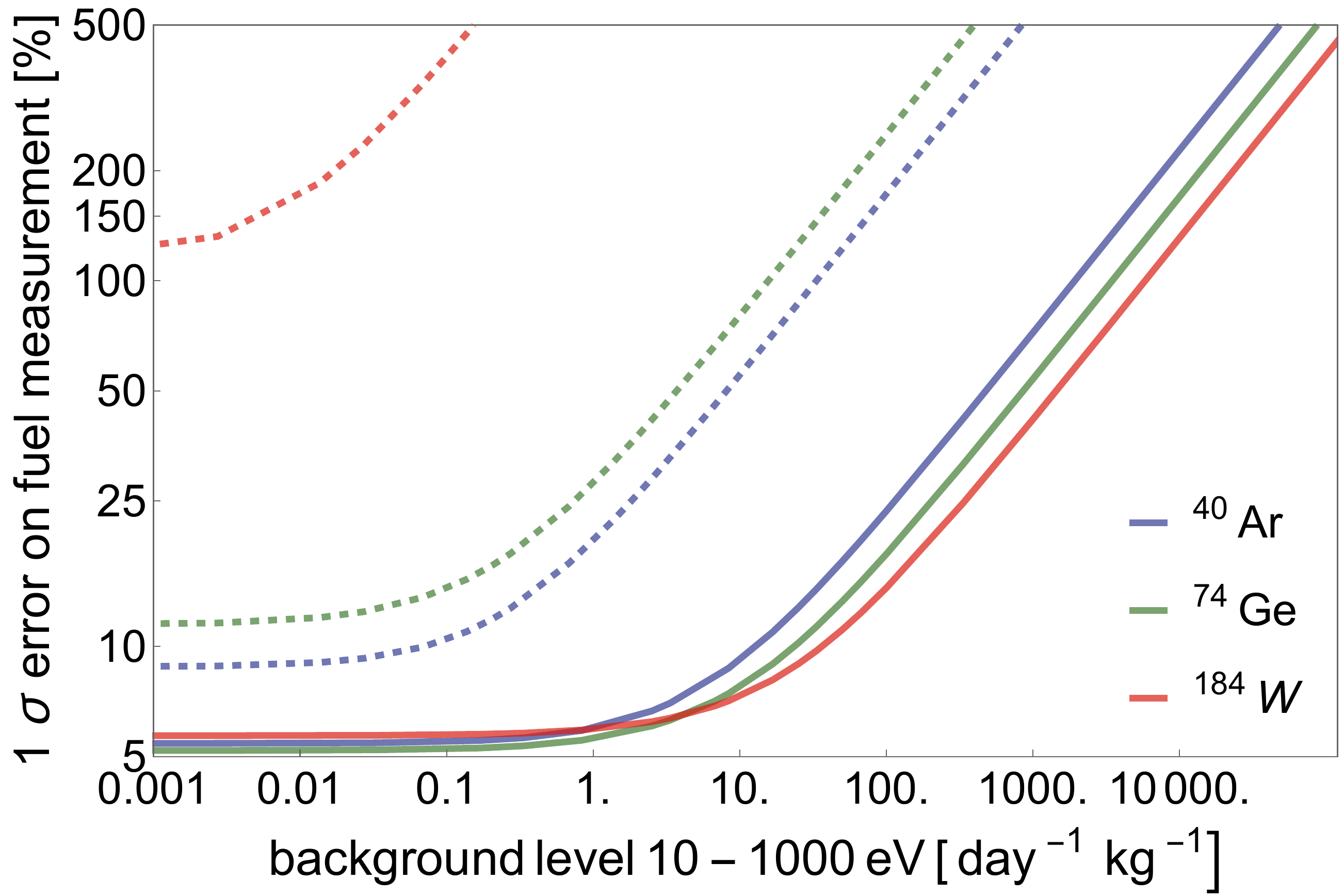}
    \caption{1 $\sigma$ error as percentages for \isotope[40]{Ar}, \isotope[74]{Ge}, and \isotope[184]{W} as a function of background levels. Solid lines indicate the 1 $\sigma$ errors of each isotope at a recoil energy threshold of 10 eV, while dotted lines of the same color indicate the 1 $\sigma$ errors at a 50 eV threshold. The detector mass is 10\,kg, the standoff is 3\,m and data taking period is one year from a 10\,MTU source.}
    \label{fig:errors}
\end{figure}
Figure~\ref{fig:errors} displays a range of backgrounds and the error percentages for each of the chosen isotopes. The isotopes \isotope[40]{Ar}, \isotope[74]{Ge}, and \isotope[184]{W} were chosen to display the ranges of measurement accuracy that could result from a detector. The minimum nuclear recoil energy resolvable was also varied from 10 to 50 eV to observe how the error increases as the resolvable energy increases.

 The error range for \isotope[184]{W} is much wider than that of \isotope[40]{Ar} or \isotope[74]{Ge}, due to the fact that the CEvNS signal from \isotope[184]{W} is strongest in the region of recoil energies between 0 and 20\,eV and effectively falls to zero around a recoil energy of 60\,eV. Thus a \isotope[184]{W} detector that can only resolve 50\,eV will be much less effective at observing a CEvNS signal. In contrast, \isotope[40]{Ar} and \isotope[74]{Ge} have much narrower error ranges, as expected. With sub-10$\%$ error levels, it is feasible that such a detector would be able to detect a single fuel element being removed amidst the $\sim10$ fuel elements in the cask. 

\section{Summary \& Outlook}
\label{sec:summary}
In this work, we present a first analysis of the potential use of a CEvNS detector to monitor spent nuclear fuel within dry storage casks, which may prove to be a unique and useful application. We have demonstrated the advantage that a CEvNS detector at the kilogram scale offers over an IBD detector for this application. Additionally, we show  that a 10 kg detector only 3 meters away from a dry storage cask over a detection time period of a year is able to detect more than 100 events for many detector isotopes at recoil energies below 100\,eV. This would allow potential sensitivity to the removal of a single fuel element form a storage casks. Backgrounds will present a significant challenge and a reduction of at least a factor $10^4$ relative to the rate for unshielded surface deployment will be required.  In terms of the criteria developed in the context of NuTools~\cite{nutools}, NuTools has clearly established criterion 1, that is a need for new technology, with its finding on spent nuclear fuel. We have demonstrated the existence of a neutrino signal (criterion 2). It seems plausible that a 10\,kg detector and a one year measurement time do not face significant implementation challenges for dry storage casks which sit in the same place for decades, thus criterion 4, implementation constraints,  appears to be met as well. This leaves, overall, criterion 3, the existence of a suitable detector technology. We encourage further investigation into the feasibility of suitable detectors for this application, which need to have a mass of around 10\,kg and a recoil threshold of less than 100\,eV combined with effective background  mitigation.

\section*{Acknowledgements}

The work was supported by the National Science Foundation REU grant
number 1757087, by the U.S. Department
of Energy Office of Science under award number DE-SC00018327 and by the 
National Nuclear Security Administration Office of Defense Nuclear
Nonproliferation R\&D through the Consortium for Monitoring,
Technology and Verification under award number DE-NA0003920. 

\bibliographystyle{apsrev-title}
\bibliography{apssamp.bib}

\end{document}